\documentclass{jetpl}

\usepackage[dvips]{graphicx}
\usepackage{bm}
\usepackage{bbm}
\usepackage{amsmath}
\usepackage{amsfonts}
\usepackage{amssymb}
\usepackage{amsthm}
\usepackage{dcolumn}

\twocolumn

\newcolumntype{.}{D{x}{}{-1}}

\newcommand{\beq}{\begin{equation}}
\newcommand{\eeq}{\end{equation}}
\newcommand{\beqn}{\begin{eqnarray}}
\newcommand{\eeqn}{\end{eqnarray}}

\newcommand{\Za}{{Z \alpha}}

\lat

\title{Two-loop self-energy contribution to the Lamb shift in H-like ions}

\rtitle{Two-loop self-energy contribution \ldots}

\sodtitle{Two-loop self-energy contribution to the Lamb shift in H-like ions}

\author{V.\,A.\,Yerokhin$^{+,*}$\thanks{e-mail: yerokhin@pcqnt1.phys.spbu.ru},
P.\,Indelicato$^{**}$, and V.\,M.\,Shabaev$^+$ }

\rauthor{Yerokhin\,V.\,A., Indelicato\,P., and Shabaev\,V.\,M.}

\sodauthor{Yerokhin, Indelicato, and Shabaev}

\address{ $^+$ Department of Physics, St. Petersburg State University,
Oulianovskaya 1, Petrodvorets, St. Petersburg 198504, Russia\\
 $^*$ Center for Advanced Studies, St. Petersburg State Polytechnical
University, Polytekhnicheskaya 29, St. Petersburg 195251, Russia\\
 $^{**}$ Laboratoire Kastler-Brossel, \'Ecole Normale Sup\'erieure et
Universit\'e P. et M. Curie, Case 74, 4 pl. Jussieu, F-75252, France }

\dates{(date to be inserted)}{*}

\abstract{The two-loop self-energy correction is evaluated to all
orders in $\Za$ for the ground-state Lamb shift of H-like ions
with $Z\geq 10$, where $Z$ is the nuclear charge number and
$\alpha$ is the fine structure constant. The results obtained are
compared with the analytical values for the $\Za$-expansion
coefficients. An extrapolation of the all-order numerical results
to $Z=1$ is presented and implications of our calculation for the
hydrogen Lamb shift are discussed.}


\begin{document}

\maketitle


Hydrogen and hydrogen-like ions are the simplest atomic systems, whose studies
inspired the creation and development of modern quantum
electrodynamics (QED). Despite their apparent simplicity, two-body atomic
systems continue to challenge physicists after more than a century of
research. On the experimental side, the absolute frequency of the $1s$-$2s$
transition in atomic hydrogen has lately been measured to 1.8 parts in
$10^{14}$ \cite{niering:00}, which represents an improvement of accuracy by 4
orders of magnitude achieved during the last twenty years. On the theoretical
side, the largest error for the $1s$-$2s$ transition energy stems presently
from the uncertainty in the experimental value for the proton charge radius
\cite{mohr:00:rmp}, even with the most recent re-analysis of
electron-proton scattering data \cite{sick:03}. Apart from that, the major
theoretical uncertainty comes from the two-loop QED effects. Important
progress achieved recently in investigations of two-loop corrections to order
$\alpha^2(\Za)^6$ \cite{pachucki:01:pra,pachucki:03:prl} resulted in a much
better theoretical understanding of the hydrogen Lamb shift. Consequently, in
the latest adjustment of fundamental constants \cite{mohr:04:rmp}, the value
for the proton charge radius (which enters as a parameter into the
determination of the Rydberg constant) was obtained by comparing
spectroscopic data for the $1s$-$2s$ transition in hydrogen with the
corresponding theoretical prediction.

The results of refs.~\cite{pachucki:01:pra,pachucki:03:prl} were obtained
within the traditional approach based on a semi-analytic expansion of binding
QED corrections in $\Za$ and $\ln [(\Za)^{-2}]$ ($Z$ is the nuclear charge
number and $\alpha$ is the fine structure constant). A peculiar feature of
the two-loop effects (first of all, the two-loop self-energy correction) is
that this expansion converges very slowly even in case of hydrogen. Numerical
values of the expansion coefficients are large and tend to grow with increase
of the order, which makes estimations of uncertainties due to
higher-order effects rather problematic. Moreover, complexity of calculations
of the $\Za$-expansion coefficients grows drastically with the increase of
their order, so that the derivation of the complete contribution to the next
order $\alpha^2(\Za)^7$ does not seem feasible in the near future.

An alternative approach is to perform the investigation
non-perturbatively in the parameter $\Za$. For one-loop
corrections, such calculations extended over three decades
\cite{mohr:74:a,soff:88:vp,blundell:91:se,jentschura:99:prl}. A
non-perturbative evaluation of two-loop QED effects is a much
more demanding problem. The dominant contribution (especially in
the low-$Z$ region) stems from the two-loop self-energy
correction, which is considered to be the most problematic
two-loop QED effect. A calculation of this correction to all
orders in $\Za$ was started in
refs.~\cite{mitrushenkov:95,mallampalli:98:pra} and completed in a
series of our investigations
\cite{yerokhin:01:sese,yerokhin:03:prl,yerokhin:03:epjd}. Up to
now, the calculation of the two-loop self-energy correction was
carried out for ions with $Z\geq 40$ only. Large numerical
cancellations growing fast when $Z$ was decreased prevented us
from calculating this correction for lower-$Z$ ions and from
drawing any definite conclusions about agreement of our results
with the known $\Za$-expansion terms. The goal of the present
investigation is to perform a calculation of the two-loop
self-energy correction for ions with $Z\geq10$ and to compare our
non-perturbative treatment with the previous investigations within
the $\Za$-expansion approach.

We start with summarizing the results available for the $\Za$ expansion of
this correction for the ground state of H-like ions. The corresponding energy
shift is conveniently represented in terms of a dimensionless function
$F(\Za)$ (relativistic units $\hbar=c=1$ are used throughout the paper), \beq
\label{FalphaZ} \Delta E  = m \left(\frac{\alpha}{\pi}\right)^2
\frac{(Z\alpha)^4}{n^3}\,F(Z\alpha)\,, \eeq where $n$ is the principal
quantum number and the $\Za$ expansion of the function $F$ reads \beqn
\label{aZexp} F(Z\alpha) &=& B_{40}+ (Z\alpha)B_{50} + (Z\alpha)^2 \Bigl[
  L^3 B_{63}
\nonumber \\ &&
  +L^2 B_{62} +  L\,B_{61} + G_{\rm h.o.}(\Za) \Bigr]
  \,,
\eeqn where  $L =\ln[(Z\alpha)^{-2}]$  and $G_{\rm h.o.}$ is the
higher-order remainder whose expansion starts with a constant,
$G_{\rm h.o.}(\Za) = B_{60}+ \Za \,(\ldots)\,$. The results
presently available for the expansion coefficients (see
refs.~\cite{mohr:00:rmp,pachucki:01:pra,pachucki:03:prl} and references 
therein) are:
\begin{eqnarray}   \label{eqB40}
 B_{40}(ns) &=& 1.409\,244\ldots\,,
 \\
 B_{50}(ns) &=& -24.2668(31)\,,
 \\
 B_{63}(ns) &=& -8/27\,,
 \\
 B_{62}(1s) &=& 16/27-(16/9) \ln 2\,,
 \\
 B_{61}(1s) &=& 49.838317 \,,
 \\     \label{eqB60}
 B_{60}(1s) &=& -61.6(9.2)\,.
\end{eqnarray}

Contrary to the calculations summarized above, our present consideration does
not rely on the $\Za$ expansion. The working frame is the Furry picture,
where the interaction of the electron with the nucleus is taken into account
to all orders right from the beginning. The price to pay is that we have to
deal with the {\em bound}-electron propagators, whose structure is much more
elaborate than that of the {\em free}-electron propagator. The detailed
analysis of the two-loop self-energy correction was presented in our previous
investigation \cite{yerokhin:03:epjd} and will not be repeated here. In this
Letter we just mention the major problems to be tackled in order to obtain
reliable numerical results in the low-$Z$ region and sketch the ways of their
solutions. (To note, numerical cancellations for the lowest value of $Z$
considered here, $Z=10$, amount to 4 orders of magnitude.)

One of the main factors that influences the numerical accuracy of
the final result is its dependence on the size of the basis
set used in the evaluation of the $P$ term (see
ref.~\cite{yerokhin:03:epjd} for details). In the present work, a
new {\it dual-kinetic-balance} \cite{shabaev:04:dkb} basis set
was employed, which considerably
improved the convergence in the low-$Z$ region. In addition, the actual
calculation was repeated several times with increasing
numbers of basis functions $N$ and then an
extrapolation $N\to\infty$ was performed. We note, however, that despite the
improvement achieved, the dependence of the $P$ term on the basis
set is still one of the major sources of the uncertainty. Another
difficulty in the evaluation of the $P$ term was to properly
control the accuracy of numerical integrations over momentum
variables. It is associated with a significant contribution coming
from the region of very large momenta, where the numerical Green
function is not smooth enough, due to restrictions of a
finite-basis-set representation. The problem has been handled by
introducing a set of subtractions that have the same behavior for
large momenta as the original integrand but are easier to evaluate
numerically, and by employing very fine grids for numerical
momentum integrations. The calculation of the $M$ term (see
ref.~\cite{yerokhin:03:epjd} for details) was carried out
employing the contour $C_{LH}$ for the integrations over the
virtual-photon energies, which is much more suitable for the
numerical evaluation in the low-$Z$ region than the integration
simply along the imaginary axis. In addition, a contribution 
containing the dominant part of the spurious behavior in the low-$Z$ region
was separated from the $M$ term and calculated
separately. It involves only one infinite partial-wave expansion, 
which makes its numerical computation easier as compared to the full $M$ term.

\begin{figure}
\centerline{
\resizebox{\columnwidth}{!}{%
  \includegraphics{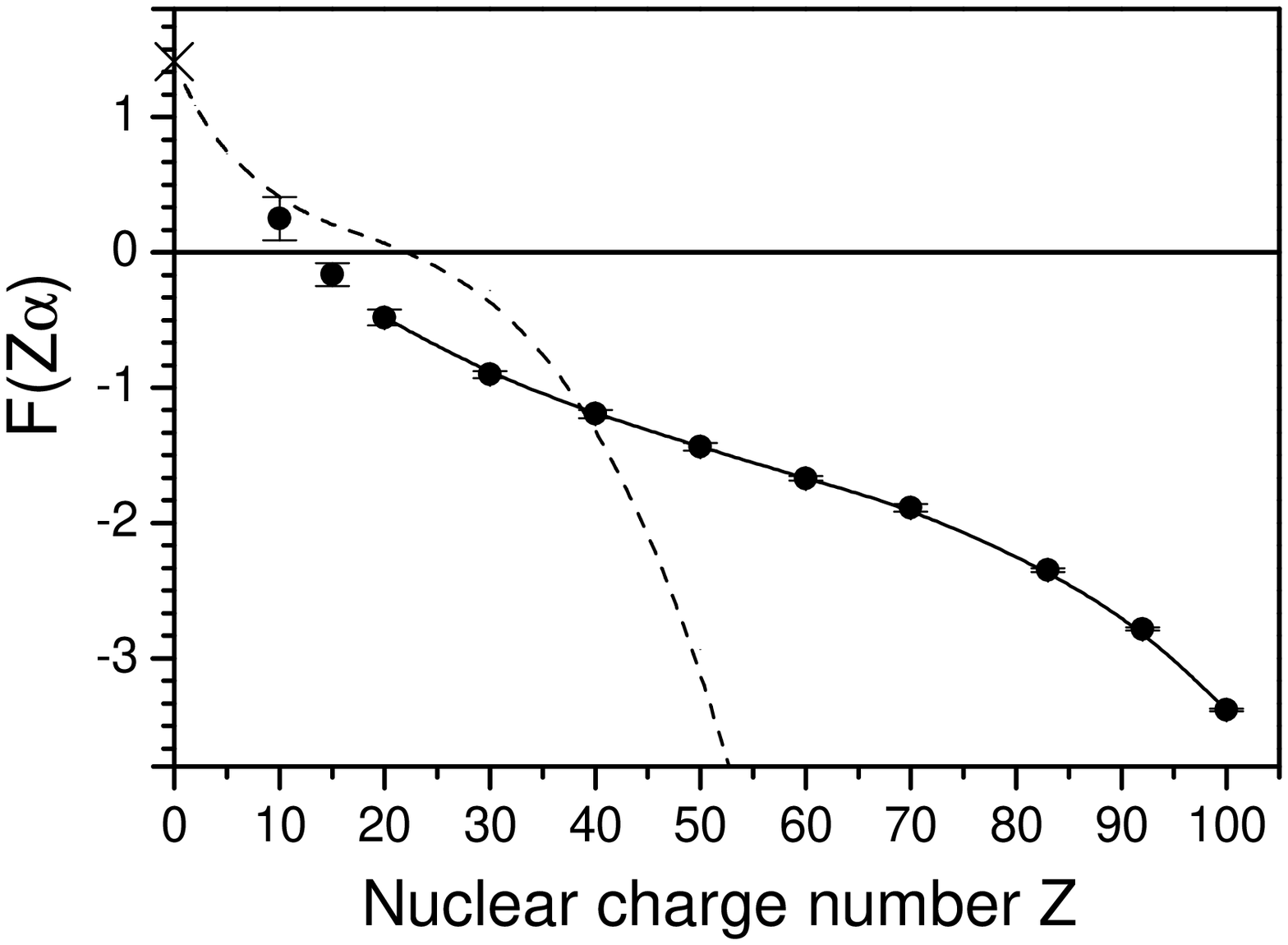}
} } \caption{The two-loop self-energy correction evaluated to all
orders in $Z\alpha$ (dots) together with the
contribution of all known terms of the $Z\alpha$ expansion (dashed
line), in terms of $F(\Za)$. The cross indicates the analytical
value of this correction at $Z=0$.
 \label{fig:FZa}}
\end{figure}

In the table, we present the numerical results obtained for the two-loop
self-energy correction for the ground state of H-like ions with $Z \geq 10$.
As compared to our previous investigations
\cite{yerokhin:03:prl,yerokhin:03:epjd}, new calculations for $Z =$ 10 -- 30
were performed and the numerical accuracy for $Z = 40$, 50, and 60 was
improved. The numerical values for $Z \geq 70$ are taken from our previous
investigation \cite{yerokhin:03:epjd}. In fig.~\ref{fig:FZa}, our
non-perturbative (in $\Za$) results are compared with the contribution of the
known coefficients of the $\Za$ expansion, as given by
eqs.~(\ref{eqB40})-(\ref{eqB60}). We observe that the numerical results
behave smoothly and tend to approach the known analytical value at $Z=0$.

\begin{figure}
\centerline{
\resizebox{\columnwidth}{!}{%
  \includegraphics{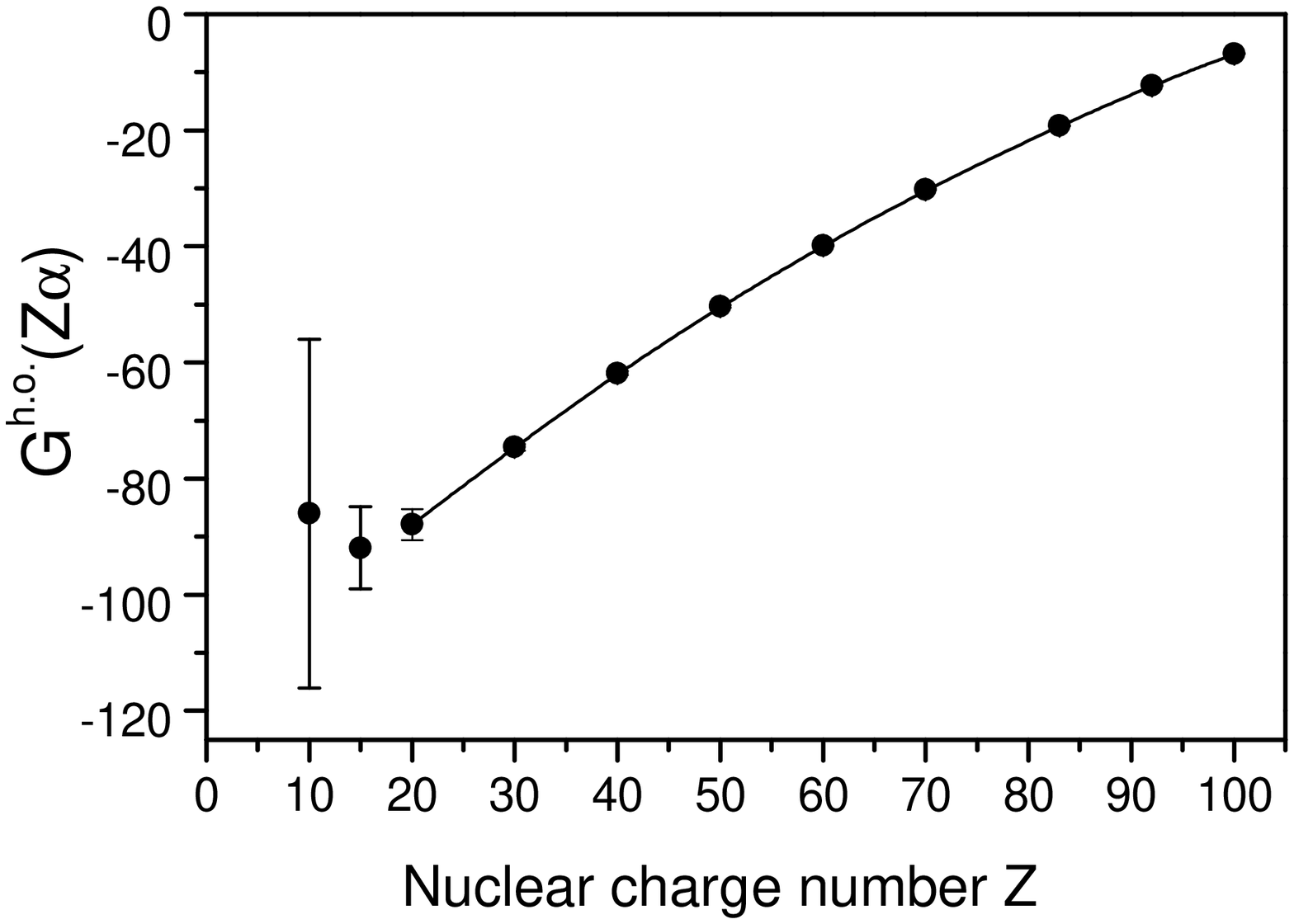}
} } \caption{Higher-order remainder $G^{\rm h.o.}(\Za)$ defined by
eq.~(\ref{aZexp}).  \label{fig:GZa}}
\end{figure}

In order to perform a more detailed comparison with the $\Za$-expansion
calculations, we separate the higher-order remainder $G^{\rm h.o.}(\Za)$
[defined by eq.~(\ref{aZexp})] from our non-perturbative results, with the
corresponding plot presented in fig.~\ref{fig:GZa}. As can be seen from the
figure, a naively estimated limit of our numerical values at $Z=0$ is about
twice as large as the analytical result (\ref{eqB60})
for the coefficient $B_{60}$. However, there is an indication
\cite{pachucki:04:priv} that the analytical result of
ref.~\cite{pachucki:01:pra} for the coefficient $B_{61}$ is incomplete. In
this case, the numerical data for $G^{\rm h.o.}$ plotted in
fig.~\ref{fig:GZa} contain an admixture of the logarithmic contribution to
the leading order and, therefore, are not bound at $Z=0$. Until analytical
calculations of $B_{61}$ have been finished, we cannot draw
any conclusions concerning agreement with the existing result for $B_{60}$.

It is possible, however, to extrapolate our numerical results for $G^{\rm
h.o.}(\Za)$ to $Z=1$ and to obtain an approximate value for the corresponding
higher-order contribution for the hydrogen Lamb shift. For such
extrapolation, we use the procedure first employed in ref.~\cite{mohr:75:prl}
and recently described in detail in ref.~\cite{bigot:03}. The approximate
value for $G^{\rm h.o.}(1\alpha)$ is obtained in two steps. First we apply an
(exact) linear fit to each pair of two consecutive points from our data set
and store the resulting value at $Z=1$ as a function of the average abscissa
of the two points of the original set. (The points with $Z=10$ and 15 are not
employed for the extrapolation because of their large error bars.) Then, we
perform a global linear or quadratic least-squares fit to the set of data
obtained and take the fitted value at $Z=1$ as a final result.

In order to test this extrapolation scheme and to check the consistency of
our numerical data with the first $\Za$-expansion coefficients, we consider
the function
\begin{equation} \label{BZa}
\widetilde{F}(\Za) = \frac{F(\Za)-B_{40}}{\Za} = B_{50}+
            (\Za)(\cdots)\,.
\end{equation}
The extrapolation procedure described above yields a result that
reproduces the analytical value for the $B_{50}$ coefficient
within the 15\% accuracy. Applying the same extrapolation scheme
to the higher-order remainder, we obtain
\begin{equation} \label{Gho}
G^{\rm h.o.}(1\alpha) = -127 \pm 30\% \,.
\end{equation}
The error bars indicated are obtained by applying the extrapolation
procedure to the function $G^{\rm h.o.}(\Za)+10\, \ln(\Za)^{-2}$, which is
supposed to account for the probable incompleteness of the present result for
the $B_{61}$ coefficient.

The result (\ref{Gho}) significantly alters the previous
prediction for the higher-order remainder \cite{pachucki:03:prl},
which reads (in our notations)
\begin{equation} \label{Ghoold}
G^{\rm h.o.}(1\alpha;\rm old) = -61.6 \pm 15\% \,.
\end{equation}
The difference between (\ref{Gho}) and (\ref{Ghoold}) leads to a shift
of the latest prediction for the hydrogen $1s$ Lamb shift
\cite{pachucki:03:prl} by 7 kHz, with the result
\begin{equation}\label{}
    \nu(1s) = 8\,172\,804\,(32)(4)\,\rm kHz\,,
\end{equation}
where the first error stems from the current uncertainty of the proton charge
radius and the second one is the theoretical uncertainty corresponding to the
one of eq.~(\ref{Gho}). We note that significant progress in the
determination of the proton radius is anticipated in the near future from the
experiment on the muonic hydrogen, which is currently being pursued in Paul
Scherrer Institute \cite{taqqu:99}.

To sum up, we have performed a non-perturbative (in $\Za$)
calculation of the two-loop self-energy correction that extends
our previous evaluation to the region $Z \geq 10$. The numerical
results obtained agree well with the first two terms of the $\Za$
expansion. A certain disagreement is found with the analytical
results to order $\alpha^2 (\Za)^6$, which could possibly be
associated with incompleteness of the present value for the
$B_{61}$ coefficient. An extrapolation of the numerical data to
$Z=1$ yields a result that alters the theoretical value for the
the hydrogen $1s$ Lamb shift by 7 kHz.

Valuable discussions with K. Pachucki and U. Jentschura are gratefully
acknowledged. This work was supported by INTAS YS grant No.~03-55-1442, by
the "Dynasty" foundation, by RFBR grant No.~04-02-17574, and by Russian
Ministry of Education (grant No.~E02-3.1-49). The computation was partly
performed on the CINES and IDRIS French national computer centers.
Laboratoire Kastler Brossel is Unit{\'e} Mixte de Recherche du CNRS
n$^{\circ}$ 8552.


%
%
\onecolumn
\begin{table}
\caption{Individual contributions to the two-loop self-energy
correction expressed in terms of $F(Z\alpha)$.  \label{tab:sese}}
\begin{center}
\begin{tabular}{r.....}
\hline\hline $Z$  &  \multicolumn{1}{c}{LAL}
              &  \multicolumn{1}{c}{$F$ term}
                             &  \multicolumn{1}{c}{$P$ term}
                                      &  \multicolumn{1}{c}{$M$ term}
                                               &  \multicolumn{1}{c}{Total}
\\ \hline
 10 &  -0.3x577  &   822.1x4(2)  &   -721.3x4(12)  & -100.1x9(10)   &     0.2x5(16)    \\
 15 &  -0.4x951  &   292.9x02(13)  &   -235.2x05(70) &  -57.3x66(48)  &    -0.1x64(85)   \\
 20 &  -0.6x015  &   136.9x11(7)   &   -102.0x26(55) &  -34.7x64(16)  &    -0.4x81(58)   \\
 30 &  -0.7x565  &    44.7x29(3)   &    -29.4x10(25) &  -15.4x65(5)   &    -0.9x03(26)   \\
 40 &  -0.8x711  &    19.5x05(3)   &    -11.5x75(30) &   -8.2x53(5)   &    -1.1x94(31)   \\
    &      x     &        x        &    -11.4x1(15)^a&   -8.2x7(18)^a &    -1.0x5(23)^a  \\
 50 &  -0.9x734  &    10.0x25(2)   &    -5.48x8(26)  &   -5.0x01(3)   &    -1.4x37(26)   \\
    &      x     &        x        &    -5.41x(8)^a  &   -4.9x9(6)^a  &    -1.3x4(10)^a  \\
 60 &  -1.0x82   &     5.7x23(1)   &    -2.97x0(18)  &   -3.3x41(2)   &    -1.6x70(18)   \\
    &      x     &        x        &    -2.93x(4)^a  &   -3.3x42(21)^a&    -1.6x3(4)^a   \\
 70 &  -1.2x16    &     3.4x97(1)  &    -1.75x7(25)  &   -2.4x12(11)  &    -1.8x88(27)  \\
 83 &  -1.4x66    &     1.9x38     &    -1.05x7(13)  &   -1.7x64(4)   &    -2.3x49(14)  \\
 92 &  -1.7x34    &     1.2x76     &    -0.81x2(10)  &   -1.5x13(3)   &    -2.7x83(10)  \\
100 &  -2.0x99    &     0.8x25     &    -0.72x3(7)   &   -1.3x84(3)   &    -3.3x81(8)   \\
\hline\hline
\end{tabular}

\vspace*{0.2cm} $^a $  ref. \cite{yerokhin:03:epjd}.
\end{center}
\end{table}

\end{document}